\documentclass{article}

\usepackage{booktabs} 
\usepackage[ruled]{algorithm2e} 

\SetAlFnt{\small}
\SetAlCapFnt{\small}
\SetAlCapNameFnt{\small}
\SetAlCapHSkip{0pt}
\IncMargin{-\parindent}
\usepackage[margin=1in]{geometry}

\usepackage[english]{babel}

\usepackage{amsmath}

\usepackage{amsmath,amsfonts,amssymb,amsthm}
\usepackage{graphicx}
\usepackage{url}
\usepackage{authblk}

\renewcommand{\proof}{{\textbf{Proof: }}}
\renewcommand{\qed}{\qedsymbol{}}

\newcommand{\score}{{\rm{score}}}

\newtheorem{theorem}{Theorem}
\newtheorem{lemma}[theorem]{Lemma}

\newtheorem{corollary}[theorem]{Corollary}
\newtheorem{observation}[theorem]{Observation}

\title{The Fairness of Redistricting Ghost}

\author[1]{Jia-Wei Liang}
\author[2]{Nina Amenta}
\affil[1]{Department of Computer Science, University of California, Davis}
\affil[2]{Department of Computer Science, University of California, Davis}

\date{}                     
\begin{document}
\maketitle

\begin{abstract}
We explore the fairness of a redistricting game
introduced by Mixon and Villar, which provides a two-party 
protocol for dividing a state into electoral districts, without
the participation of an independent authority. 
We analyze the game in an abstract 
setting that ignores the geographic distribution of voters
and assumes that voter preferences are fixed and known. 
We show that
the minority player can always win at least $p-1$ districts, 
where $p$ is proportional to the percentage of minority voters. 
We give an upper bound on the number of districts won by the minority based on  
a ``cracking" strategy for the majority. 
\end{abstract}


\section{Introduction}

In 1974 Montana was the first US state to adopt an 
independent redistricting commission for the House of 
Representatives, consisting of two Republicans, two Democrats, and
an independent chair. 
During the most recent redistricting cycle, 
some of the proposed maps splitting the state into two districts
concentrated the Democrats into one competitive district,
while others consisted of two reliably 
Republican districts. 
The choice came down to the judgement of one person, the independent chair \cite{Montana}.

Independent commissions have been 
adopted as an alternative
to redistricting by state legislatures, which have an obvious partisan bias. 
But redistricting commissions also struggle to establish a perception of fairness. 
They suffer from the 
appearance or existence of bias, a lack of transparency, and difficulty coming 
to an agreement;
some have failed to produce maps at all, while others have fallen into legal 
quagmires \cite{aba,propublica}. 
Similarly when the courts are used to challenge 
redistricting plans, again 
the decision falls to a jury, a panel of judges, or a single judge, whose fairness can be questioned. 

We can imagine many other possible mechanisms, 
or protocols, for 
redistricting. 
If we could find and adopt a redistricting 
protocol that was broadly 
seen as fair by voters, this could contribute to the 
perceived legitimacy of democratic systems. 
While this would be novel, so were independent commissions 
before 1974.  

Only a small fraction of the
mathematical research into redistricting has focused
on fair protocols;
instead the focus is often on the immediately useful topic of
defining when 
a proposed map is unfair. 
One approach is to define fairness properties which 
the map, or an election on the map, should have.
Another is to define distributions of acceptable maps 
in order to identify a proposed 
map as an outlier. 
A fair protocol might 
be useful in these endeavors as well. 
It could lend validity to the
properties that its redistricting solutions exhibit, 
and 
simulation of a fair protocol could be used to generate distributions
of maps, which would also inherit legitimacy 
from the protocol. 

In this paper, we study a proposed
protocol,
due to Mixon and Villar \cite{mixon2018utility}, related to 
well-known mechanisms for allocating 
players to sports teams. 
The reader may recall the ``captains" method from schoolyard games:  
two team captains are chosen (somehow), and then the captains take turns 
choosing players. 
The ``snake draft" used to assign players to 
teams in professional sports (eg. the NFL draft or Fantasy Football) is a similar turn-taking mechanism. 
This approach is familiar to voters, 
and perceived as fair, at least in the
realm of sports. 

The analogy between
redistricting and choosing sports teams is 
incomplete in several ways. 
The two parties in redistricting construct a map with several 
districts, not two teams. 
Also, there are constraints (differing from state to state) 
on what constitutes a valid map. 
Fundamentally, all districts must contain equal numbers of voters.  
There are also restrictions on the 
connectivity and shape of the districts, 
constraints ensuring adequate racial representation, and
goals (or soft constraints) such as alignment with municipal or other local boundaries. 

\textbf{Redistricting Ghost:} 
Mixon and Villar's protocol handles many of these issues. 
We call their protocol
Redistricting Ghost, because, as they describe,
the game is inspired by the word game Ghost.
In Ghost,
two players take turns adding letters to a string, and a player loses if they 
are the first to spell a word.  
Each player has to 
demonstrate, after their turn, that the string they have constructed is a prefix of an English word. 
In Redistricting Ghost, two players 
$A$ and $B$, representing the two parties, take turns assigning 
a voter to a district (instead of 
individual voters, the game could also be played with 
pre-selected equal-sized groups of voters, for example census tracts).  
On his or her turn, a player places any voter 
into any district. 
After their turn, the player must be able to display a complete valid map,
meeting whatever legal constraints there are, 
that extends the current set of partially defined districts. 
Thus Redistricting Ghost accommodates 
legal constraints on maps and also resembles the 
mechanisms for picking sports teams. 

\textbf{Our results:} 
Redistricting Ghost may ``feel" fair, but what can we say mathematically?
Mixon and Villar proved Theorem~\ref{thm:mirror}, below, which handles 
the special case of a perfectly tied electorate in an abstract setting. 
We keep the abstract setting but consider arbitrarily sized majorities and numbers of districts.  
We find that Redistricting Ghost is 
reasonably fair from the perspective of the minority 
party $B$. 
In particular, let 
$p$ be the number of districts that $B$ should win to 
be proportional to the size of their minority. 
We show that the minority party 
has a strategy that can always win at least $p-1$ districts.

We also consider a ``cracking" strategy for the majority party; that is, a strategy 
that attempts to distribute the minority voters uniformly over all the districts. 
This classic strategy is essential to gerrymandering, and, when the majority 
has complete control over redistricting, leads to the majority winning every district. 
As a strategy in Redistricting Ghost, however, we find that 
cracking is not very strong; 
it limits the minority to at most $p-1$ districts only when 
the minority is very small. 

\section{Related Literature}
\label{sec:priorWork}

Redistricting, as an important element of representative democracy,
is of great interest in the law, political science and economics, 
the press, and recently mathematics and computer science.
Much of the mathematical literature concerns detecting or 
quantifying gerrymandering in a given allocation of voters to districts,
including metrics such as the efficiency gap \cite{efficiencygap} 
and statistical analysis of distributions of possible maps. 
This paper is most closely related to the 
smaller body of research on
protocols by which political parties can negotiate or collaboratively determine
an electoral map. 

We build on the analysis of 
Redistricting Ghost in \cite{mixon2018utility}
considered an abstract non-geometric setting  
in which everyone can perfectly predict the party each voter will vote for, 
and there are no geographic, geometric, 
demographic or other constraints on assigning voters to districts. 
In this abstract setting they 
proved 
\begin{theorem}
\label{thm:mirror}
(Mixon and Villar) 
Let $j$, the number of districts, be even,
and let both parties have the same number 
$n = v/2$ of voters. 
If they play optimally, then both players
win exactly $j/2$ districts. 
\end{theorem}
The proof of this theorem is a ``mirroring" strategy 
for the second player,
arbitrarily $B$. 
Before the game starts, 
$B$ matches each district with 
another, its ``mirror".
$B$ then responds to each move by $A$ with a mirror move. 
For example, when $A$ places one of her 
voters into a district, 
$B$ counters by placing one of his voters into   
its mirror district. 
Similarly when $A$ places one of $B$'s voters into a district, $B$ places one of $A$'s voters into its mirror district. 
In the end $A$ and $B$ win the same number
of districts.
This mirroring strategy requires $j$ to be even. 
Also, it does $B$ no good when he is the minority, 
since $A$ can successfully take a cracking approach. 
In most situations, we need a different analysis.

Most other research into protocols has addressed  
cake-cutting mechanisms for redistricting, which generalize the 
well-know ``I cut, you choose" protocol for splitting a cake in two. 
Like sports drafts, cake-cutting can accommodate
arbitrary, possibly different, objective functions for the two players.
And like Redistricting Ghost, 
cake-cutting can also incorporate legal map constraints. 

Landau et al. \cite{fair1,fair2}
proposed an approach in which an independent authority creates a set of nested initial cuts, and 
then parties $A$ and $B$ follow a protocol to choose one of the cuts, and 
to assign one side to each of the parties. 
Each party may then gerrymander their side as they see fit. 
If the protocol fails, then the sides are assigned randomly. 
They prove (Theorem 6.1) a \textit{Good Choice Property}: that when the nested cuts are chosen fairly by the independent authority, 
both parties will achieve a result near the average of the 
best and worst possible results (based on their 
individual objective functions) of any map 
respecting the chosen cut. 
This mechanism involves an independent authority and randomization, both of which we
would like to avoid. 

Pegden, Procaccia and Yu \cite{cut} proposed the 
\textit{I-cut-you-freeze} protocol, a game in which the two
parties switch roles at every turn, with each turn adding one ``frozen" district to 
the map.  
For example, during her turn $A$ extends 
the existing set of frozen districts to a complete map, drawing new districts in the so-far empty part of the 
state subject to the legal constraints. 
Then $B$ selects one of the newly drawn districts to freeze;
the rest of the extended map is discarded. 
They give a tight bound (Theorem 2.4) on the number of
districts that either player can win in the abstract setting,  
which shows that the number of districts won 
by either player is close to proportional. 
The majority player $A$ does a bit 
better than proportional representation when the minority is small, 
but (as with Redistricting Ghost) $B$ can always win at least $p-1$ districts. 
In addition, they show that even in the abstract setting their protocol prevents 
any designated protected population from being packed into a single district. 
A drawback is that the party that goes first has a significant advantage when the 
number of districts is small, 
which does not seem fair.

Recently Ludden et al. \cite{ludden2022bisection} proposed a bisection 
protocol, in which $A$ and $B$ take turns bisecting every remaining large-enough 
district into two smaller districts, of equal size up to rounding. 
They give a symmetric optimal strategy for both players,
extensive analysis using
simulations comparing bisection and I-cut-you-freeze, and 
some analysis in a semi-geometric (graph-based) setting.
In the abstract setting,
while they do not
completely characterize the maps produced by the bisection 
protocol 
they do prove some properties.  
One result (Lemma 1) is that when $j$ is a power of two, 
the minority player requires 
an $\Omega(1/\sqrt{j})$ fraction of the voters to win one district; this 
suggests that the majority player has a significant advantage. 
Also, again, the player who goes first has a significant 
advantage when the number of districts is small. 

Tucker-Foltz \cite{cutAndChoose} proposed a 
more theoretical game in which $A$ is allowed to redistrict
as they please, but then $B$ picks a threshold for the election. 
Any district 
where the margin of victory does not meet the threshold is assigned randomly to 
either party with equal probability. 
He showed that in the Nash equilibria for this game the expected 
number of districts 
won by each player differs from proportional representation 
by at most one. 
The completely packed map (defined in Section~\ref{sec:prelim})
is one of these Nash equilibria, and all equilibria require 
some packed majority districts, unless the number of voters is equal. 
Because of its heavy use of randomization, this 
protocol is unlikely to be seen as fair by voters.

\section{Definitions and Notation}

Following \cite{mixon2018utility}, we define Redistricting Ghost on a state
with two parties, $A$ and $B$. 
The parties use colors (a)pple green and (b)rick red, 
respectively, and we'll call their voters apples and bricks. 
Working in the abstract setting, 
we assume that every voter is
consistently an apple or consistently a brick, 
and that players $A$ and $B$ know which are which. 
We'll assume there are more apples than bricks, 
so that $B$ is the minority party.  

In Redistricting Ghost, 
$A$ and $B$ take turns, with $B$ going first.  
At each move, a player adds a voter to any district which is not yet full. 
Player $A$ can play either an apple or a brick, and similarly $B$.

\begin{figure}[htb]
    \centering
    \includegraphics[scale=0.7]{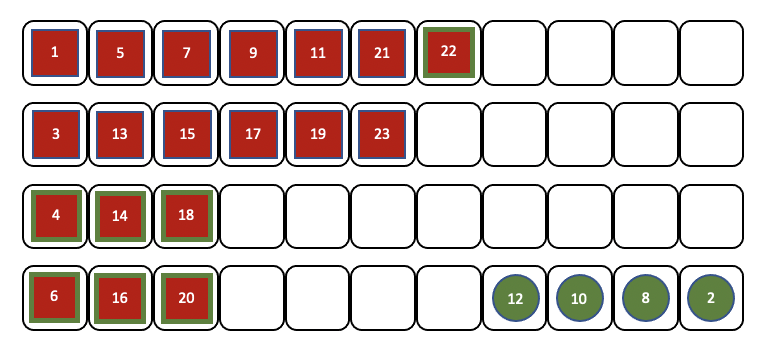}
    \caption{
    A simple example of Redistricting Ghost. An interactive version of the game can be found at \emph{http://ballsandbins.com}.
    Each row represents a district; so here $j = 4$. 
    The size of a district is $2m+1$, $m=5$ here.
    We draw bricks as red squares and apples as green circles; a brick played by $A$ has a green outline, and an
    apple played by $B$ has a red outline. The number inside each brick or apple represents the turn
    at which it was played ($B$ plays odd turns and $A$ plays even).  
    For visual clarity, we place bricks into districts from left to right and
    apples from right to left, and we sort the rows by the number of bricks they contain, and within that by the 
    number of empty spaces. 
    Here, we see the state of the game after the last brick has been played; at every subsequent
    turn an apple will be played to an empty position, and the order 
    in which they are played will have no effect on the outcome. 
    So $n$, the number of bricks, is $19$, and $q$, the number of districts won by $B$, is $2$.}
    
    \label{fig:example}
\end{figure}

There are $j$ districts, 
each with positions for $2m+1$ voters, 
so that the total number of voters 
$v = j(2m+1)$. 
Let $n$ be the total number of bricks, and
let $q$ be the number of districts won by $B$ at the end of the game. 
Figure~\ref{fig:example} shows an example of game play and illustrates the notation.

If a district $d$ contains more apples than bricks, we say 
$A$ is ahead in $d$, or that $d$ is an ``apple district", and similarly $B$. 
If $d$ contains equal numbers of apples and bricks then
$d$ is tied (this occurs during the game but not at the end). 
If there are at least $m+1$ bricks $d$ at the end of the game, 
we say $B$ wins the district, and similarly $A$.
An ``open district" is one which contains fewer than $2m+1$ voters.

A \textit{strategy for $B$} (or respectively $A$) is an algorithm describing how 
$B$ should play at every turn. 
The strategy for $B$ may reference how $A$ has played in 
earlier turns, but it does not 
assume that $A$ plays a particular strategy, 
and visa versa. 

In our analysis, we take $j,q$ and $m$ as given and characterize how many bricks
$n$ are necessary or sufficient for $B$ to win $q$ districts.
In realistic situations $m$ is orders of magnitude larger
than $j$ and $q$, 
so we will sometimes 
ignore additive constants that depend only on $j$ and $q$, since they make little difference
as $m \rightarrow \infty$.
We do not consider bounds that are 
asymptotic in $j$, the number of districts; 
$j$ is always much smaller than $m$. 

One way in which the abstract setting is misleading 
is that we defined it so that a district is never tied; when a district contains
$m+1$ bricks and $m$ apples, say, we count it as won by $B$. 
In reality a nearly-even district could go either way in an election, or 
could even be exactly tied.
We will see later that this artifact produces some skirmishing early in the game, 
but the overall strategies of the players, when $m$ is large, 
do not seem to depend on this property. 
Also, while we do not prove that the results 
of Redistricting Ghost are indifferent to which 
player makes the first move, it does not 
seem to be an important factor. 

\section{Some Fairness Properties}
\label{sec:prelim}

After analyzing
Redistricting Ghost, we
will compare its results to other measures of the fairness
of a map. 

At the end of the game, we define 
$m+1$ of the bricks in a brick district, and 
$m+1$ of the apples in an apple district, to be \textit{useful} 
(they were needed to win the district) and 
the rest of the voters in the district to be \textit{wasted}. 
Since $m$ voters in each district are wasted, the total number of wasted votes
is always $v * m/(2m+1)$; but one party's votes may be wasted more than the other's. 
The \textit{efficiency gap} \cite{efficiencygap} 
is the difference $E$ in the number of wasted voters for each party 
as a fraction of the total number of voters.
So $0 \leq E \leq 1/2$; and a 
large efficiency gap - say greater than $1/4$ - is taken as a sign of gerrymandering.
In the case of Montana, 
a reliably Republican map has an efficiency gap near zero, while depending on the result
of an election a map containing a swing district might have an efficiency gap of either zero, 
if the Republicans win, or $1/2$ if the Democrats win.  The reliably Republican map seems 
more fair, then, when judged by the efficiency gap. 

We define the 
proportional representation for the minority party as 
$$
p = \mbox{round}(jn/v)
$$
where $n$ is the number of bricks, $j$ is the number of districts, $v$ is the
total number of voters, and $\mbox{round}$ is the function that
rounds up or down to the nearest integer.
In the abstract setting, there is a deterministic assignment of voters to districts - the ``packed map" - 
that achieves the proportional representation, as follows. 
We completely fill as many districts 
as we can with bricks, completely fill as many districts as we can with apples, 
and finally construct at most one mixed
district containing both apples and bricks.  
Then $p$ is the number of districts $B$ wins using this allocation. 
In Montana, a map containing a contested district is similar to the packed map, 
and thus more likely to provide proportional representation in an election; 
so judging by proportional representation, a map with a contested district seems more fair. 

Since $B$ is the minority party, $p \leq \lfloor j/2 \rfloor$. 
There is a range of $n$ 
corresponding to each value of $p$:
\begin{lemma}
\label{lem:rangeOfP}
For any value of $p$, we have 
$$
p + (2p-1)m \leq n \leq p + (2p+1)m
$$
\label{lem:pBounds}
\end{lemma}
Proof: 
$B$ barely wins 
$p$ districts in the packed map
with $p-1$ districts packed with bricks and 
$m+1$ bricks in the single mixed district. 
And $B$ will win only $p$ districts when 
there are $p$ districts packed with bricks
and $m$ bricks in the mixed district. 
Thus
$$
(p-1)(2m+1) + m+1 \leq n \leq p(2m+1) + m 
$$
Simplifying gives us the bound. 

\section{Strategy for the Minority Player}

Let $q$ be the number of districts that $B$ is guaranteed to be able to win using this strategy. 
Using $q$, we define a score at any point in the game, 
which will measure how well $B$ is 
doing in their quest to win $q$ districts. 
The score of district $d$ is defined to be $m+1$ if it contains at least 
$m+1$ bricks, zero if it contains at least $m+1$ apples, 
and the number of bricks in $d$ otherwise. 
If there is a set $Q$ of $q$ districts in which $B$ is either ahead or tied (possibly including empty districts), the score of $Q$ is 
$$
\score(Q) = \sum_{d \in Q} \score(d)
$$
The score of the game is the 
maximum score of any choice of $Q$.  
When there is no set of $q$ districts in which 
$B$ is ahead or tied, we say the score is zero. 
%
A set $Q$ achieving the maximum score is 
a \textit{maximizing} $Q$. 
At the beginning of the game, the score is zero, 
and, if $B$ succeeds in winning $q$ districts, at the end
of the game the score is $q(m+1)$.  
Define $u$ to be the minimum score
of any district in a maximizing $Q$. 
\begin{lemma} 
The minimum score $u$ 
is the same for any maximizing $Q$, and the number of districts in $Q$ 
with $\score(d) = u$ is the same for any maximizing $Q$.
\end{lemma}
Proof: Assume for the purpose of contradiction that
some maximizing $Q_1$ 
has more districts with minimum $\score(d) = u$ 
than some other maximizing $Q_2$.  
Then we could replace some district $d_1$ in $Q_1$  with 
$\score(d)=u$ with a district $d_2$ from $Q_2$ with 
$\score(d_2) > u$, 
raising the score of $Q_1$.  
This contradicts the assumption that $Q_1$ is maximizing. 
\qed
\begin{corollary}
Every maximizing $Q$ contains the same number of empty 
districts.
\end{corollary}
\begin{observation}
Consider any maximizing district $Q$.
Every district $d$ not in $Q$ either 
is an apple district or has $\score(d) \leq u$.
\label{obs:lowPotential}
\end{observation}

Let $b$ be the number of moves 
by $B$ 
that increase $B$'s score, 
and let $h$ (for ``helping") be the number
of moves by $A$ that increase $B$'s score. 
Both kinds of moves 
necessarily involve playing a brick.  
Let $w$ be the number of moves by $A$ that waste a brick, 
that is, $A$ plays a brick but does
not increase $B$'s score. 

\begin{lemma}
\label{lem:enoughBricks}
Any strategy by which $B$ can
increase the score by at least one at each of his turns, so long as any bricks remain to be played,
will allow $B$ to win $q$ districts if $n \geq 2(q(m+1) - h)$; 
since $h \geq 0$, this means that
$B$ can always win $q$ districts if $n \geq 2q(m+1)$. 
\end{lemma}
\proof
Define a round to consist of a move by $A$, followed by a move by $B$. 
If the score 
increases at $q(m+1)$ or more turns,
then $B$ will win $q$ districts. 
We have
$$
b \geq w + h
$$
since, in every round, $B$ plays a brick and $A$ either helps 
with a brick, wastes a brick, or plays an apple. 
We also have 
$$
n = 1+(b-1)+h+w = b+h+w \leq 2b 
$$
since $B$ plays a brick to start off the game, and then $B$ can play a brick to improve the score in each round, and $A$ might play a brick in each round. 
Now assume we have enough bricks, as defined in the statement of the Lemma, so that
$$
2 b \geq n \geq 2(q(m+1) - h).
$$
We simplify this to 
$$
b + h \geq q(m+1)
$$
which implies that $B$ has won $q$ districts. 
\qed
\\[1ex]

Next, we define a strategy for the minority player $B$, in 
Algorithm~\ref{alg:minority}. 
We will argue that with this strategy $B$ 
can indeed increase the score
in each round.
\begin{algorithm}[thb]
\SetAlgoNoLine
\If{ no bricks remain}
    {Play an apple to any open district\;
    break\;}
{Select a maxmizing $Q$ including the fewest (possibly zero) tied districts. } \;
\If {$Q$ contains a non-empty tied district $d$} 
            {\tcp{Type a move}
            Play a brick to $d$\; 
            break\;}
\ElseIf{$Q$ contains at least one empty district $d$} 
            {\tcp{Type b move}
            Play a brick to $d$\;
            break\;}
\Else
            {\tcp{$Q$ contains no tied districts}
            \tcp{Type c move}
            Play a brick to any district $d$ in $Q$ containing $\leq m+1$ bricks.} 
\caption{Strategy for the Minority Player}
\label{alg:minority}
\end{algorithm}

Our argument that $B$ will be able to increase his score at every round is based on:
\begin{lemma}
At the beginning of a round, fix a maximizing $Q$, and 
let $z$ be the total number of empty districts (in or outside of $Q$).
Assume that: 
\\1. $Q$ exists, 
\\2. $Q$ includes brick 
districts and empty districts, but no 
non-empty tied districts, and 
\\3. The number of empty districts in $Q$ is 
at most $\lfloor z/2 \rfloor$.
\\
If $B$ plays the strategy of Algorithm 1, 
and a brick remains for him to play at his turn, 
then these three conditions 
will continue to hold at the beginning of the next round. 
\label{lem:induction}
\end{lemma}
\proof
Recall that a round is a move by $A$ followed by a move by $B$. 
We divide $A$'s possible moves into two categories:
$A$ might play to an occupied district, 
or $A$ might play to an empty district (in or out of $Q$). 

First, assume $A$ plays to an occupied district $d$. 
If $A$ plays an apple to an apple district or a brick to a
brick district, the conditions still hold.
If $A$ plays an apple to a brick district $d$ it might become tied.
If $d \not \in Q$, the conditions still hold. 
If $d \in Q$, $B$ then makes a move of type $a$, restoring Condition 2. 
Finally, if $A$ plays a
brick to an apple district $d$ it might become tied. 
If it becomes part of every maximizing $Q$, 
again, $B$ makes a move of type $a$, restoring Condition 2. 

Next, we consider the case that $A$ 
places a brick in an empty district $d$. 
Then $B$ makes a move of type $b$ or $c$.
The number of empty districts $z$ decreases by one, 
and, if $Q$ contained any empty districts before $A$'s move, 
it is replaced by $d$,
the number of empty districts in $Q$ goes down by one, and Condition 3 still 
holds. 

Finally, assume $A$ places an apple in an empty district $d$.  
If there are no empty districts in $Q$, then Condition 3 still holds. 
If $d \not\in Q$,  
and there is an empty district $d'$ in $Q$, 
$B$ will play a brick to $d'$ (a move of type $b$),
restoring Condition 3. 
Finally if $d \in Q$, 
then $d$ drops out of $Q$, but
Condition $3$ implies that there is at least one other empty 
district $d'$ which replaces $d$ in $Q$, and again 
$B$ make a move of type $b$, restoring Condition 3.
\qed

\begin{theorem} 
If $n \geq 2q(m+1)$, playing the strategy in Algorithm 1 ensures that $B$ will
win at least $q$ districts.
\label{thm:upperBound}
\end{theorem} 
\proof 
We use induction on the number of rounds. 
At the beginning of the game, the three conditions of Lemma~\ref{lem:induction} hold.
So as long 
as $B$ plays using the strategy of Algorithm 1, the three conditions of Lemma~\ref{lem:induction} 
will continue to hold at the next round.  
Following the strategy, as long as there are remaining bricks, $B$ makes a moves of type a, b or c. 
All of these add a brick to an existing maximizing $Q$, increasing the score by one. 
Thus $B$ increases the score in every round, and
Lemma~\ref{lem:enoughBricks} thus ensures that $B$ wins
$q$ districts.
\qed

\section {Strategy for the Majority Player}

Now we want to show that the majority player 
$A$ can prevent $B$ from winning $q$ districts
when $n$ is too small; that is,
we will get a lower bound on the number of
bricks required for $B$ to win $q$ districts, as a function 
of $j$ and $m$.  
In particular, we will show
\begin{theorem}
The minority player $B$ can win $q$ districts only if
$$
n \geq f(q) = 2q \left( 1 - \frac{q}{j+q} \right) (m + 1) - 1 
$$
\label{thm:lower}
\end{theorem}
\begin{figure}[htb]
    \centering
    \includegraphics[width=0.7\textwidth]{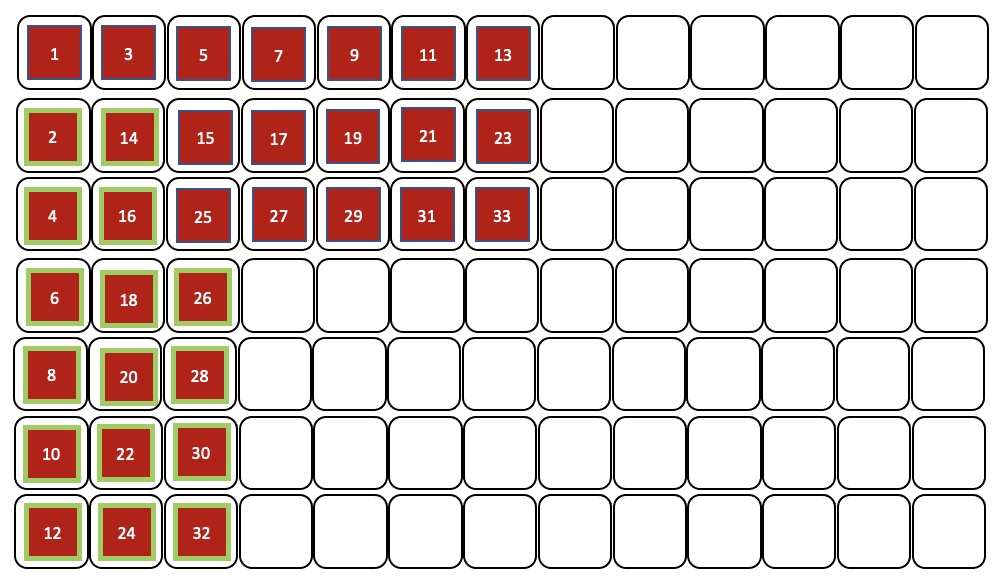}
    \caption{An example of a game in which the majority player $A$ uses the ``cracking" 
    strategy in Algorithm~\ref{alg:majorityNew}, at the point at which
    the bricks run out. 
    The main idea of of the algorithm is that $A$ 
    fills in as many columns as she can
    with bricks. Player $B$ may do anything.
    In this example, $B$ concentrates on winning one district at a 
    time.
    The lower bound (Theorem~\ref{thm:lower})
    says that if the number of bricks $n < 28$, 
    then $B$ cannot win three districts.  Here $n = 33$, and $B$ just barely wins three districts.  With $q=3, j=7, m=6$, 
    we prove that $A$ can always fill $c=2$ columns, 
    but in this example $A$ actually manages to fill $3$ columns. 
    }
    \label{fig:majorityExample1}
\end{figure}

\begin{algorithm}[bht]
\SetAlgoNoLine
{Given $n$, find the largest value of $q$ such that
$n < f(q)$, and use $q$ to find $c$}\;
{In each round:}\;
\If{ no bricks remain}
     {Play an apple to any open district\;}
\ElseIf{ there is any open district where the number of open spaces $< 2(c-r)$, where $r$ is the number of bricks in the district}
    {\tcp{Type a move}
    Play a brick to that district \;}
\Else
    {\tcp{Type b move}
    Let $i$ be the least number of bricks in any open district\;
    Play a brick to an open district containing $i$ bricks\;}
\caption{Strategy for the Majority Player}
\label{alg:majorityNew}
\end{algorithm}

\begin{figure}[hbt]
    \centering
    \includegraphics[width=0.7\textwidth]{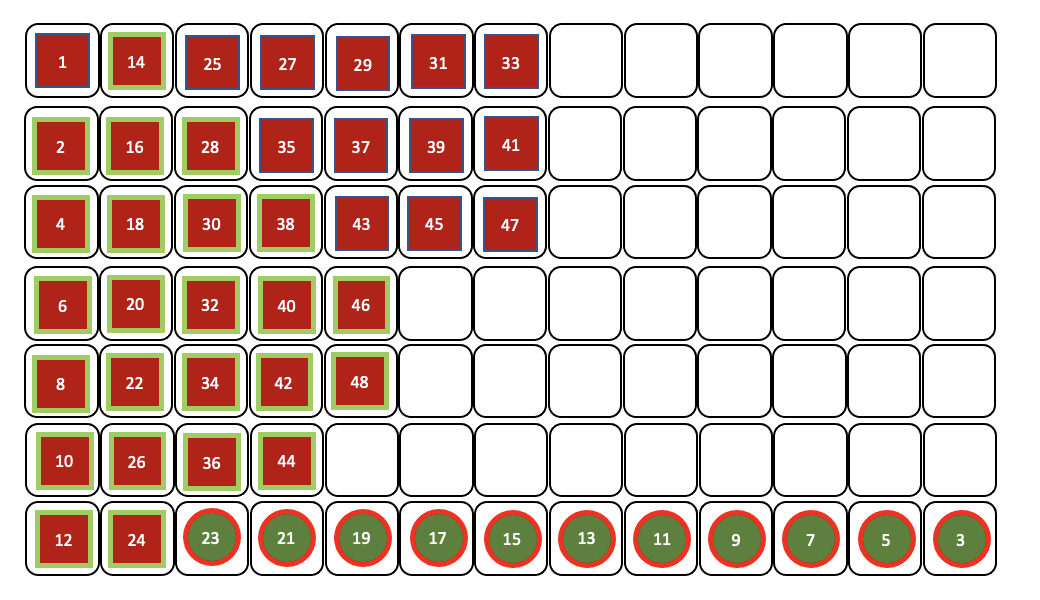}
    \caption{ Another example game in which $A$ plays the ``cracking" strategy, at the point at which the bricks run out. 
    In this game $B$ fills in a row with apples so 
    as to block $A$ from adding the final brick to columns 3 and 4, and then switches to winning districts one by one.  
    Using the notation of
    Theorem~\ref{thm:lower}, we have
    $j=7, m=6, q=3, c=2$, and $n=37$.
    We notice that $B$ needs $n=36$ to win three districts, while he only needed $33$ in Figure~\ref{fig:majorityExample1}.
    }
    \label{fig:majorityExample2}
\end{figure}

We will prove Theorem~\ref{thm:lower} using the strategy for 
$A$ that appears in Algorithm \ref{alg:majorityNew}.
In this strategy 
$A$ uses Theorem~\ref{thm:lower} to choose
the smallest value of $q$ to which they can limit 
$B$.
Using $q$, $A$ plays a classic ``cracking" strategy
in which they ensure that at least 
$$
c = \left \lfloor \frac{q}{j+q} (m+1) - \frac{1}{j+q} \right \rfloor
$$ 
columns are filled with bricks at the end of the game. 
The Type a moves (see the algorithm) keep 
the first $c$ columns free of apples; this makes the analysis easier. 
Recall that an ``open district" is one which contains fewer than $2m+1$ voters.

This strategy makes no sense unless $f(q)$ is
monotone in $q$, so that a larger $q$ requires a 
larger $n$. 
Fortunately,
\begin{observation}
If $q_1 > q_0$, then $f(q_1) > f(q_0)$.
\end{observation}
It also requires the following
\begin{observation}
In any round, 
there is at most one district to which $A$ can make
a move of type a. 
\end{observation}
This is because a move of type a is 
triggered by $B$ placing an 
apple into a district, reducing the number of 
open spaces $\omega$ to $2(c-r)-1$. 
The Type a move increases $r$, restoring 
$\omega = 2(c-r)$.
This gives us
\begin{lemma}
If  $A$ does not run out of bricks, 
there will be at least $c$ bricks in 
each district at the end of the game. 
\label{lem:fillC}
\end{lemma}
Now that we have established that the strategy makes
sense, let's proceed to 
\\[1ex]
\textbf{Proof of Theorem~\ref{thm:lower}:}
Assume at the end of the game that $B$ has won 
$q$ districts.  

The upper-left 
rectangle of size $q \times (m+1)$ contains
only bricks.  
We claim that at least 
the first $c$ columns 
also contain only bricks. 
So assume for the purpose of contradiction that at most
$c-1$ columns are filled with bricks. 

In this case, we claim that every 
move by $A$ placed a brick into 
the first $c$ columns. 
This is always true for moves of type $a$, 
and because the first $c$ columns are not 
full, and cannot contain apples, it will 
be true of moves of type $b$ as well. 

This means that all of the bricks outside of the 
first $c$ columns - of which there are at least $q (m+1-c)$ - 
must have been placed by $B$.  
To each of these moves, except possibly the last, 
$A$ responded by placing 
a brick into the first $c$ columns. 
It takes no more than $j$ bricks to fill a column, 
so it must be that 
$cj$ is strictly larger than the number of bricks in the first $c$ columns
\begin{eqnarray*}
cj &>& q(m+1-c) - 1 \\
c (j+q) &>& q(m+1) + 1 \\
c &>& \frac {q(m+1)}{j+q} - \frac{1}{j+q}
\end{eqnarray*}
This contradicts the definition of $c$, 
so it must be the 
case that the first $c$ columns are filled with 
bricks. 

So the total number of bricks must be 
\begin{eqnarray*}
 n &\geq& q(m+1) + (j-q)c  \\
 &\geq& q(m+1) + (j-q) \left \lfloor \frac{q}{j+q} (m+1) - \frac{1}{j+q} \right \rfloor \\
 &\geq& q(m+1) + \frac{q (j-q)}{j+q} (m+1) - 1 \\
 &\geq& q(m+1) + q(m+1) - \frac{2 q^2}{j+q} (m+1) - 1 \\
 &\geq&  2q \left( 1 - \frac{q}{j+q} \right) (m + 1) - 1
\end{eqnarray*}\qed

\section{Fairness Relative to Proportional Allocation}

Theorem~\ref{thm:lower} describes the 
values of $n$ below which $B$ cannot win $q$ districts, and Theorem~\ref{thm:upperBound} 
describes the values of $n$ above which $B$ can always win at least $q$ districts.
As a sanity check, we note that
the range of $n$ where we do not know either that $B$ can or that 
$B$ cannot win $q$ districts is
$$
2q(m+1) \geq n \geq 2q \left( 1 - \frac{q}{j+q} \right) (m + 1) - 1 
$$
and we see that this gap always exists.

Next, we recall that
the proportional outcome is for $B$ to win $p$ districts, and that this
implies that the number of bricks $n$ lies in a specific range:
$$
(2p-1) m + p \leq n \leq (2p+1) m + p. 
$$
We consider the breakpoint values of $n$ at which $p$ changes. 
At the smallest value of $n$ at which the proportional representation 
is $p$, 
Theorem~\ref{thm:upperBound} tells us 
that $B$ can always win at least $p-1$ districts:
$$
n = p + (2p-1)m \geq 2(p-1)(m+1)
$$
Theorem~\ref{thm:lower} tells us that $B$ cannot win $p$ districts when $p$ is small:
\begin{eqnarray*}
n = p + (2p-1)m &\leq& 2p \left( 1 - \frac{p}{j+p} \right) (m+1) - 1 \\
2p(m+1) - m - p &\leq& 2p(m+1) - \frac{2p^2}{j+p} m - \frac{p}{j+p} - 1 \\
m+p &\geq& \frac{2p^2}{j+p} m + \frac{p}{j+p} + 1
\end{eqnarray*}
for instance, when $p < \sqrt{j}$. 


\section{Discussion}
\label{sec:discussion}

\begin{figure}[tbh]  
   \centering
   \includegraphics[scale=0.5]{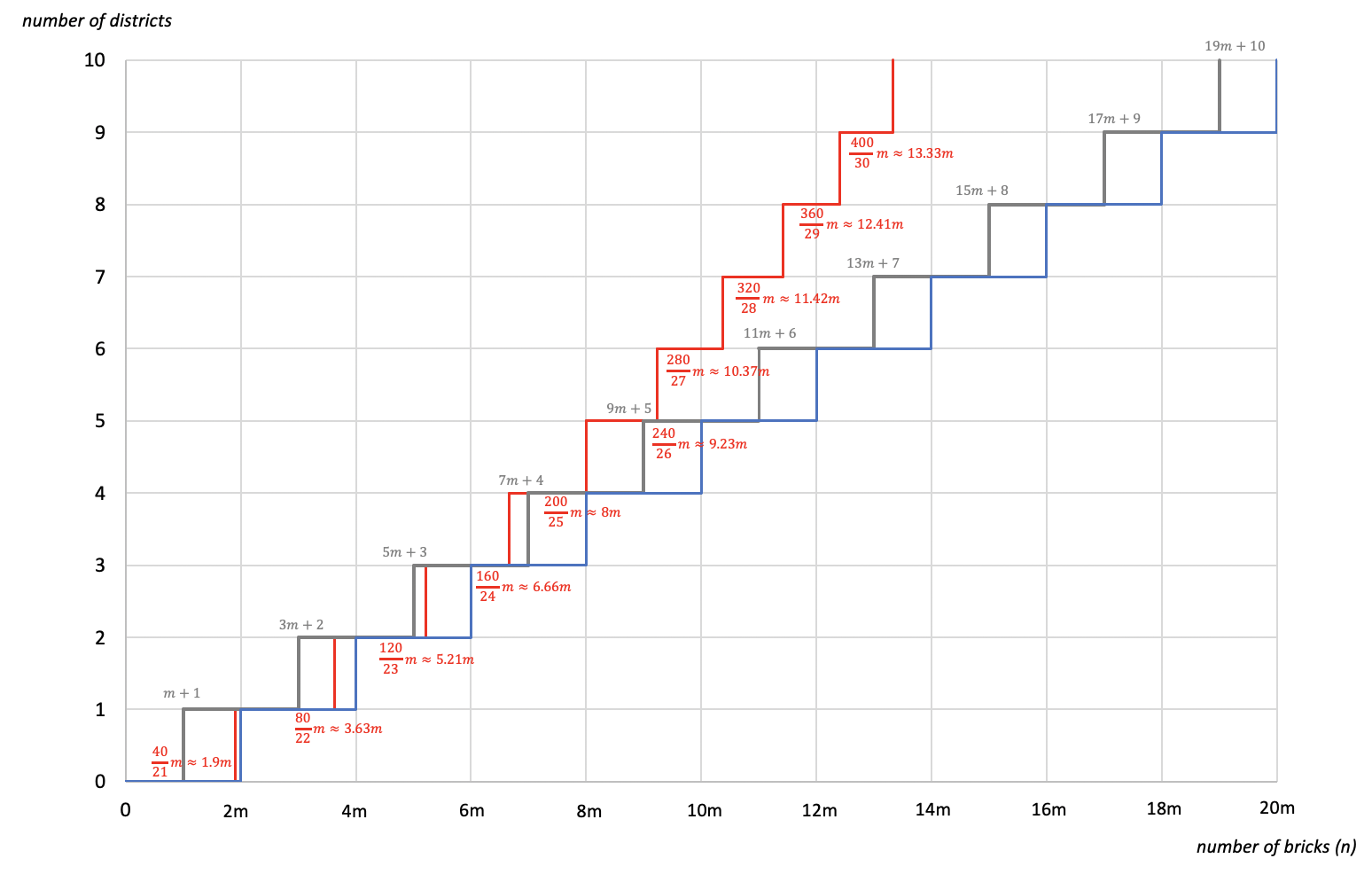}
   \caption{The bounds illustrated for the case $j=10$.  
   The $x$ axis is the number of bricks (voters of the minority party $B$) 
   as a function 
   of $m$, the size of a bare majority in a district. The $y$ axis 
   is the number of districts that $B$ can win. 
   The grey plot shows $p$, the number of districts $B$ 
   wins if they are distributed proportionally. 
   The red plot shows $n=\frac{2jq}{j+q}m$; 
   Theorem~\ref{thm:lower} says that
   when $n$ is below this value, $B$ cannot win $q$ districts. 
   The blue line shows $n=2qm$; Theorem~\ref{thm:upperBound} says that 
   when $n$ is above this value, $B$ can always win $q$ districts. }
 \label{fig:graph}
\end{figure}

To visually 
compare the majority and minority players' strategies, 
we graph an example for a game of reasonable size ($j=10$) in Figure~\ref{fig:graph}. 
%
%
%
We see that the gap between the lower bound (blue line) and upper bound (red line) on the number of minority voters $n$ required to win $q$ districts 
increases with $q$. 
We conjecture that improving the strategy for the minority player $B$
would show that $B$ can win more than $p$ districts when $n$ is large (although of course never more than $j/2$). 
That is, we conjecture that Redistricting Ghost favors the minority player when the
minority is large, 
and the majority player when the minority is small. 
We further conjecture that these results depend only trivially on which player goes first.

It will be important to understand the results of Redistricting Ghost in more realistic settings, involving geometric and legal constraints on the districts.
A good first step might be to analyze the protocol in a model that 
simplifies the distribution and geometry with 
a graph, as in \cite{deford2019recombination}. 

Redistricting Ghost \cite{mixon2018utility}, like I-cut-you-freeze \cite{cut} and the
bisection protocol \cite{ludden2022bisection}, does not allow the minority player to achieve proportional
representation in the abstract setting when the minority is small.  
The fact that several protocols show the same effect suggests that it might
be an inherent
feature of the redistricting problem: there are many 
ways to ``crack" a small group of minority voters and prevent them from dominating
any one district.
If this is inherent to the problem, 
one could consider this to be fair, that is, that small minorities should
not expect to achieve proportional representation via redistricting, 
or perhaps it shows that the 
idea of electing representatives using districts, even independent of the geographic distribution of voters, is inherently unfair. 



\section{Acknowledgements}
\label{sec:acknowlegements}

We thank an anonymous reviewer for pointing out an error in an earlier version of this paper. 
We thank (funding sources temporarily omitted) for support during this project.

\bibliographystyle{ACM-Reference-Format}
\bibliography{sample}
\end{document}